\documentclass{llncs}
\usepackage[utf8]{inputenc}
\usepackage[T1]{fontenc}

\usepackage{cite}
\usepackage{mathtools}
\usepackage{enumerate}

\usepackage[pdfpagelabels=true]{hyperref}
\usepackage{cleveref}
\usepackage{xcolor}
\usepackage{algorithm2e}
\usepackage{csquotes}
\usepackage{url}

\newcounter{algorithm}

\newif\iflongversion

\hypersetup{
	linktoc = page,
	pdfpagemode = UseNone,
	colorlinks,
	linkcolor={red!50!black},
	citecolor={blue!50!black},
	urlcolor={blue!80!black}
}

\usepackage{paralist}
\usepackage{ragged2e}
\usepackage{tikz}
\usepackage{pgfplots}
\usepackage{tikz-qtree}
\usepackage[linguistics]{forest}
\usetikzlibrary{arrows}
\usetikzlibrary{backgrounds, decorations.pathreplacing, bending}

\newcommand*{\NEG}{\textcolor{blue}{\texttt{$-$}}\xspace}
\newcommand*{\POS}{\textcolor{blue}{\texttt{$+$}}\xspace}

\begin{document}

    \title{Near-Optimal Pool Testing under Urgency Constraints}

    \author{\'Eric Brier\inst{1} \and Megi Dervishi\inst{2} \and Rémi Géraud-Stewart\inst{2,3} \and David Naccache\inst{2}
    \and Ofer~Yifrach-Stav\inst{2} 
    } 

    \institute{
	\email{eric.brier@polytechnique.org }\\
    \and
    DI\'ENS, \'ENS, CNRS, PSL University, Paris, France\\
45 rue d'Ulm, 75230, Paris \textsc{cedex} 05, France\\
\email{\url{ofer.friedman@ens.fr}}, 
\email{\url{given\_name.family\_name@ens.fr}}
    \and 
    Qualcomm Inc., San Diego, USA \\
    \email{rgerauds@qti.qualcomm.com}
}

\maketitle

\begin{abstract}
Detection of rare traits or diseases in a large population is challenging. Pool testing allows covering larger swathes of population at a reduced cost, while simplifying logistics. However, testing precision decreases as it becomes unclear which member of a pool made the global test positive.\smallskip

In this paper we discuss testing strategies that provably approach best-possible strategy --- optimal in the sense that no other strategy can give exact results with fewer tests. Our algorithms guarantee that they provide a complete and exact result for every individual, without exceeding $1/0.99$ times the number of tests the optimal strategy would require.

\smallskip 

This threshold is arbitrary: algorithms closer to the optimal bound can be described, however their complexity increases, making them less practical. 

Moreover, the way the algorithms process input samples leads to some individuals’ status to be known sooner, thus allowing to take urgency into account when assigning individuals to tests.

\keywords{Pool Testing, Probability, Information  Theory, Optimality, Adaptive and Non-Adaptive Testing Strategies}
\end{abstract}
 
\section{Introduction}

In this paper, we discuss pool testing (also known as group testing) strategies from an information-theoretic perspective.
From this point of view, any single test tells us something about the status --- positive or negative --- of individuals in a population. Since there is no way to learn in this fashion more information than there is, a natural figure of merit for a given testing strategy is to measure what proportion of the total information it collects.

We shall say throughout this paper that a strategy is \emph{near-optimal} when this proportion (better described as the ratio of the average number of tests to the entropy of the tested population) exceeds 99\,\%. Note that this threshold is arbitrary and serves to give a concrete instantiation of our methods, yielding relatively simple strategies.

The rest of this paper describes testing strategies that are near-optimal in a range of simplified, but realistic situations. 

Our aim is to describe these strategies in a way that is immediately applicable to real-world situations, such as detection of SARS-COV-2. Their near-optimality is easy to check, and we introduce a compact graphical notation for them. 

Near-optimality does not necessarily imply optimality; however by definition no strategy can outperform ours by more than 1\,\% in terms of number of tests performed. 

The mathematical methods and theory that enabled us to design these testing strategies are highly non-trivial, and we defer their complete description to another paper.

\subsection{Pool Testing and Shannon Entropy}

The connection between pool testing and information theory was first made by Sobel and Groll in 1959 \cite{sobel1959group,aldridge2019group}. We recall here their argument for the sake of clarity.

We consider a population, with each individual being either \emph{positive} ($+$) or \emph{negative} ($-$). We do not assume anything about what this labeling means medically. However we consider that it is possible to \emph{pool-test} a group of individuals: by \enquote{mixing together} their samples, and testing the resulting mix, we obtain a certain outcome ($+$ or $-$). If any of the samples from this pool was $+$, then the outcome is $+$. Alternatively, if the pool-test outcome is $-$, then no individual from the tested group was $+$.

This method is well-known and practical (see \Cref{sec:medical}), within technical and ethical limits which are not within the scope of this paper. We assume that the tests have negligible error rates. We also do not take into account dilution effects (i.e. the fact that the greater the pool is, the greater the chance is for false negative).

If the total population consists of $n$ individuals, each carrying one information bit (whether they are $+$ or $-$), then there is an $n$-bit string $S$ describing the status of every individual. Testing one individual reveals the corresponding bit of $S$. Naturally, testing all individuals one by one reveals the complete string $S$. Trivially, any binary test (such as pool testing) reveals, again, at most one bit of information.

Shannon's entropy measures $H(S)$, the amount of bits necessary to describe $S$. Therefore, any testing strategy providing complete and correct information on $S$ must perform, on average, at least $H(S)$ tests. An \enquote{optimal} testing strategy would perform no more than $H(S)$ tests.
A near-optimal strategy approaches this situation arbitrarily closely, within a ratio of $1 - \epsilon$. In this paper we chose $\epsilon = 0.01$ to keep the exposition simple and concrete.

Advanced testing strategies better approaching the optimum exist, but their description is more intricate and would only result in marginal practical advantages over the strategies described in this paper. With that in mind, researchers interested in applying such strategies to real-world scenari are strongly encouraged to contact the authors.

Finally, adaptive pool testing in the presence of a large percentage
of positives is best done by individual testing, rather than by pooling. However, the positiveness probability making individual testing optimal is not known with certainty.

\subsection{Context and Related Work} \label{sec:medical}

\paragraph{Pool testing in theory.}
Pool testing was first formally studied by Dorfman in 1943 \cite{Dorfman}, who suggested using it to detect syphilis in the US military. In Dorfman's approach, pools of identical sizes are formed, and positive pools are retested one by one. Using pools of size $n$, for a homogeneous population of $N$ individuals and a positive probability $p$, Dorfman's method performs on average
\begin{equation*}
    \frac{N}{n}(1 + n(1 - (1-p)^n))
\end{equation*}
tests. This can be inverted to yield the optimal pool size $n^\star$, which maximizes the number of tested individuals. Dorfman shows that
\begin{equation*}
    n^\star = \frac{2}{\ln(1-p)}W\left(- \frac12 \sqrt{-\ln(1-p)} \right)
\end{equation*}
where $W$ is the Lambert $W$ function\footnote{This functions is defined as follows for any complex number $z$:
$z = w =e^w \; \Longleftrightarrow \; w = W(z)$.
}. Following Dorfman, many variants and improvements were suggested \cite{morris2006overview}: Sterret \cite{sterrett1957detection}, halving methods \cite{litvak1994screening}; some extensions which can leverage a priori knowledge of some heterogeneity in the population \cite{mcmahan2012informative,bilder2010informative,black2012group}; and combinatorial algorithms \cite{li1962sequential,du2000combinatorial,beunardeau2020optimal}.

Hwang's generalised binary-splitting algorithm (1972) \cite{hwang1972method} works by performing a binary search on groups that test positive, and is a simple algorithm that finds a single defective in no more than the information-theoretic lower-bound number of tests. This has been improved by Allemann in 2013, with an algorithm performing $0.255d+ \frac12 \log_{2}(d)+5.5$ tests above the information lower bound when $n/d\geq 38$ and $d\geq 10$, where $d$ is the quantity of positive individuals \cite{allemann2013efficient}.

All the testing strategies discussed so far are \emph{adaptive}, in the sense that they may retest individuals based on the result of previous tests. The search for efficient and near-optimal \emph{non-adaptive} tests is still a very open problem, motivated by the desire to perform tests in parallel and at scale, and can also be approached from an information-theoretic angle \cite{coja2019information}.

\subsubsection{Pool Testing in Practice.} Besides syphilis \cite{Dorfman}, pool testing has been used
in the detection of influenza \cite{van2012pooling}, chlamydia \cite{currie2004pooling}, malaria \cite{taylor2010high}, HIV \cite{emmanuel1988pooling}, and more recently, SARS-CoV-2 \cite{hogan2020sample,yelin2020evaluation,sinnott2020evaluation,shani2020efficient,torres2020pooling,eberhardt2020multi}.

The latter has received an intense interest due to the pandemic's fast expansion, uncertainty about prophylactic measures, absence of efficient treatment, atop the threat on lives and hospital capacity. Testing remains to this day the only way to catch carriers of SARS-CoV-2 at an early stage, which greatly increases the hope of limiting contagion, as well as successful recovery for the individual \cite{who:2020}.

While relatively efficient and precise tests were quickly developed, producing them at scale and distributing them is more of an issue. Test shortages \cite{erdman:2020} and financial constraints made it necessary to reduce the costs associated with mass testing: many countries including Germany \cite{liu2020pool}, Israel \cite{BenAmi2020}, Korea \cite{Korea:2020}, the United States \cite{Nebraska:2020}, and India \cite{India:2020} have adopted pool testing as their \emph{de facto} standard. 
However, pool testing has practical limitations that make its applicability sub-optimal: dilution while pooling makes detection in large pools difficult\footnote{For SARS-CoV-2, RT-PCR tests can work with pools of size about 32 \cite{yelin2020evaluation}, which is far beyond what currently done in practice, around 5 to 10 \cite{hogan2020sample}.}; retesting individuals may be difficult, impossible or undesirable; the construction of the mixtures, which is done by technicians by hand, can be time-consuming and error-prone; error rates of actual tests may be sensitive to the marker's concentration, and pooling may cause the result to be unexploitable due to large error margins.

\section{Preliminaries}

The following subsections will describe a set of testing strategies, called \emph{algorithms} and will provide a high-level summary of the resulting performances. All mathematical computations providing performance estimates are deferred to the appendices of this paper. 

\subsection{Graphical Representation of Algorithms}

To describe the proposed procedures without ambiguity, we adopt the following graphical representation:
\begin{itemize}
    \item Each algorithm is represented as a tree read from left to right. Each node has two branches, top and bottom, whose precise meaning is described below. 
    \item Letters at the edges (e.g., $A$, $B$, etc.) stand for individuals being pool-tested together at each testing step.
    \item 
Leaves indicate samples that are determined \emph{negative} (denoted \NEG) or \emph{positive} (denoted \POS). We write $(\POS, \NEG, \NEG, \POS, \dotsc)$ to mean that $A$ is \POS, $B$ and $C$ are \NEG, $D$ is \POS etc.  

    \item 
A bar over a letter (e.g., $\overline{A}$) denotes \emph{introduction}, namely the operation consisting in randomly drawing a new individual from the queue and assigning to it the concerned letter (e.g. $\overline A$ means: \enquote{draw a random individual from the queue and denote it by $A$}).
See \Cref{fig:unit}.

\begin{figure}[!htp]
    \centering 
    \begin{tikzpicture}[grow=right,level distance=2cm]
\Tree[
.\node{$\overline{A}$} ;
	\edge[green] ;[.\node[blue]{\texttt{($-$)}} ; ] 
	\edge[red] ; [.\node[blue]{\texttt{($+$)}} ; ] 
]
\end{tikzpicture}
    \caption{A unary test: draw a random individual from the queue and test it.}\label{fig:unit}
\end{figure}
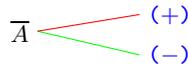
\item 
When more than one population is sampled, the use of uppercase and lowercase letters is used to distinguish between the two populations.
\item 
Branches to the bottom represent negative results, and are marked in green. Branches to the top represent positive results, and are marked in red. For instance, \Cref{fig:unit} shows the classical test of one patient.
\item 
A leaf labeled with the letter \textcolor{blue}{\texttt{R}} means that the concerned individual ought to be \enquote{recycled} (or \emph{re-pooled}) in a subsequent test. The reader may be surprised that we re-pool, and may be worried that, in doing so, we somehow lose information. However, this is not the case: as we will detail further below, since in fact, no information was learnt about this patient.

\item 
Finally, edges can carry orange labels (e.g., $\mbox{\textcolor{orange}{\texttt{L4}}:~}\overline{A},\overline{B},\overline{C}$). This allows jumping to the concerned edge and repeating a tree branch again. Note that labels are always associated with barred letters (redraw and resume). 
\end{itemize}

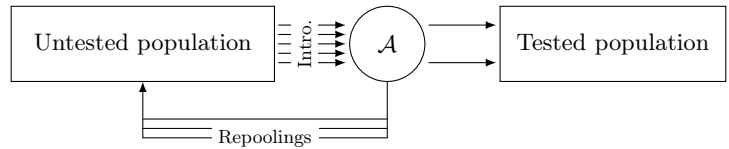
\begin{figure}[!ht]
    \centering
    \begin{tikzpicture}
    \draw (0, 0) rectangle (3.5, 1); \node at (1.75, 0.5) {Untested population};
    \draw (6.5, 0) rectangle (9.5, 1); \node at (8, 0.5) {Tested population};
    \draw (5, 0.5) circle (0.5cm); \node at (5, 0.5) {$\mathcal A$};
    \draw[>=latex,->] (3.55, 0.5) to (4.45, 0.5);
    \draw[>=latex,->] (3.55, 0.375) to (4.45, 0.375);
    \draw[>=latex,->] (3.55, 0.25) to (4.45, 0.25);
    \draw[>=latex,->] (3.55, 0.75) to (4.45, 0.75);
    \draw[>=latex,->] (3.55, 0.625) to (4.45, 0.625);
    \draw[>=latex,->] (5.55, 0.25) to (6.45, 0.25);
    \draw[>=latex,->] (5.55, 0.75) to (6.45, 0.75);
    \node[rotate=90,fill=white,scale=0.8] at (3.9, 0.5) {Intro.};
    \draw (1.75, -0.5) to (5, -0.5);
    \draw (1.75, -0.625) to (5, -0.625);
    \draw[>=latex,->] (5, 0) to (5, -0.75) to node[midway,scale=0.8,fill=white]{Repoolings} (1.75, -0.75) to (1.75, 0);
    \end{tikzpicture}
    \caption{A high-level overview of a step during a population test, using one of our algorithms denoted here $\mathcal A$.}
    \label{fig:overview}
\end{figure}

\begin{remark}
The number of individuals being tested (because of introduction), as well as the number of results obtained out of our algorithms (because of re-pooling), depend on successive test results. Thus our algorithms are best interpreted as \enquote{streaming} tests that progressively consume an untested population and produce individual test results. As mentioned earlier, \enquote{untested} is to be understood in an information-theoretical sense, and is therefore equivalent to stating that we know nothing of its test result. This operation is illustrated in \Cref{fig:overview}.
\end{remark}

\begin{remark}
It may happen in practice that one or several \emph{introductions} fail, due to the lack of available untested individuals. This can only happen when the remaining untested population is small, which in practical terms means at most a couple of times. When that happens, we can skip the introduction or equivalently we can draw an already-tested,  known-to-be-negative individual. The final result is unaffected, as is the total number of tests performed.  
\end{remark}

\subsection{Dealing with Urgency Constraints}
Unlike other pool-testing strategies where a positive result in a pool yields no information about the individuals consisting the pool, this method allows to guarantee that certain individuals in the pool will get a result. We can, therefore, predict for which position(s) in the testing scheme results are guaranteed. This allows to prioritize individuals within the testing process without delaying the results of other individuals or adding more load to the system.

For example, in Algorithm $\mathcal A_3$ below, the individuals $C$ and $E$ are guaranteed to receive a testing result (negative or positive), whereas other individuals contribute information to the pool, but themselves may be returned to the group of individuals awaiting the test, and will be tested again with another pool. In Algorithm $\mathcal A_4$, for example, it is individual $D$ who is guaranteed to get testing results.

\subsection{Homogeneous and Non-homogeneous Populations}

We assume prior knowledge of a risk level, in the form of a probability $x$ that an individual tests positive. Several models can be considered:
\begin{itemize}
    \item In the \emph{homogeneous population} model, $x$ is the same for every individual;
    \item In the \emph{non-homogeneous population} model, $x$ depends on the individual being considered (e.g. weight);
    \item In the \emph{stratified population} model, the population is divided into subgroups, which are assumed to be homogeneous (e.g. age group). 
\end{itemize}
Depending on the model and on the values of $x$, certain strategies are better than others. In a first time, we focus on the homogeneous model, providing a set of algorithms that achieve above 99\% optimality in a large range of values of $x$. Then we address the stratified model where we show how to combine the aforementioned algorithms to achieve again at least 99\% optimality in a large range of values of $x$.


\section{Homogeneous Population Algorithms}

The algorithms in this section perform tests in an homogeneous population. We first describe \enquote{basic} algorithms, which are then used to generate an infinite family of \enquote{compound} algorithms. Finally, we discuss the ranges of probability $x$ over which these algorithms achieve 99\% optimality.

\subsection{Basic Algorithms}
\subsubsection{Algorithm $\mathcal A_1$.}
This algorithm consists in the unary test of a single individual.

\subsubsection{Algorithm $\mathcal A_2$.}
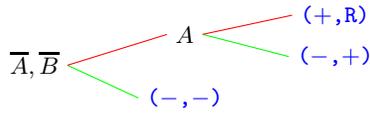
\begin{figure}[!htp]
    \centering
    \begin{tikzpicture}[grow=right,level distance=2cm]
\Tree[
.\node{$\overline{A},\overline{B}$} ;
	\edge[green] ; [.\node[blue]{\texttt{($-$,$-$)}} ; ] 
	\edge[red] ; [.\node{$A$} ;
       	\edge[green] ; [.\node[blue]{\texttt{($-$,$+$)}} ; ]
       	\edge[red] ; [.\node[blue]{\texttt{($+$,R)}} ; ]
        ]
	]
]
\end{tikzpicture}%
    \caption{Algorithm $\mathcal A_2$.}\label{facto2}
\end{figure}

This algorithm performs a pairwise test with re-pooling, see \Cref{facto2}.

\subsubsection{Algorithm $\mathcal A_3$.}
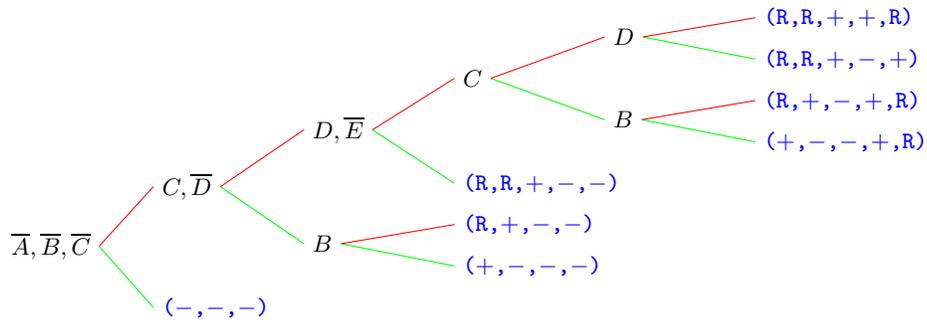
\begin{figure}[!htp]
    \centering
    \begin{tikzpicture}[grow=right,level distance=2cm]
\tikzset{every tree node/.style={anchor=base west}}
\Tree
[.\node{$\overline{A},\overline{B},\overline{C}$};
    \edge[green] ; 
        [.\node[blue]{\texttt{($-$,$-$,$-$)}} ; ] 
	\edge[red] ; 
	    [.\node{$C,\overline{D}$} ;
       	    \edge[green] ; 
       	    [.\node{\texttt{$B$}} ;
       	        \edge[green] ; 
       	            [.\node[blue]{\texttt{($+$,$-$,$-$,$-$)}} ; ] 
       	        \edge[red] ; 
       	            [.\node[blue]{\texttt{(R,$+$,$-$,$-$)}} ; ]
       	    ]
       	    \edge[red] ; 
       	        [.\node{\texttt{$D,\overline{E}$}} ; 
       	            \edge[green] ; 
       	                [.\node[blue]{\texttt{(R,R,$+$,$-$,$-$)}} ; ]
       	            \edge[red] ; 
       	                [.\node{\texttt{$C$}} ; 
       	                    \edge[green] ; 
       	                        [.\node{\texttt{$B$}} ; 
       	                            \edge[green] ; 
       	                                [.\node[blue]{\texttt{($+$,$-$,$-$,$+$,R)}} ; ]
       	                            \edge[red] ; 
       	                                [.\node[blue]{\texttt{(R,$+$,$-$,$+$,R)}} ; ]
       	                        ]
       	                    \edge[red] ; 
       	                        [.\node{\texttt{$D$}} ; 
       	                            \edge[green] ; 
       	                                [.\node[blue]{\texttt{(R,R,$+$,$-$,$+$)}} ; ]
       	                            \edge[red] ; 
       	                                [.\node[blue]{\texttt{(R,R,$+$,$+$,R)}} ; ]
       	                        ]
               	        ]
                ]
	   ]
]
\end{tikzpicture}%
    \caption{Algorithm $\mathcal A_3$.}\label{facto3}
\end{figure}

This algorithm performs an initial three-wise test, with subsequent introductions and re-pooling, see \Cref{facto3}.

\subsubsection{Algorithm $\mathcal A_4$.}
\begin{figure}[!htp]
    \centering
    \begin{tikzpicture}[grow=right,level distance=2cm]
\tikzset{every tree node/.style={anchor=base west}}
\Tree[
.\node{$\bar{A},\bar{B},\bar{C},\bar{D}$} ;
	\edge[green] ; [.\node[blue]{\texttt{($-$,$-$,$-$,$-$)}} ; ] 
	\edge[red] ; [.\node{$C,D$} ;
       	\edge[green] ; [.\node{\texttt{$B$}} ;
       	    \edge[green] ; [.\node[blue]{\texttt{($+$,$-$,$-$,$-$)}} ; ]
       	    \edge[red] ; [.\node[blue]{\texttt{(R,$+$,$-$,$-$)}} ; ]
       	    ]
       	 \edge[red] ; [.\node{\texttt{$D$}} ; 
       	    \edge[green] ; [.\node[blue]{\texttt{(R,R,$+$,$-$)}} ; ]
       	    \edge[red] ; [.\node[blue]{\texttt{(R,R,R,$+$)}} ; ]
             ]
        ]
	]
]
\end{tikzpicture}%
    \caption{Algorithm $\mathcal A_4$. Note that D is never re-pooled.}\label{facto4}
\end{figure}
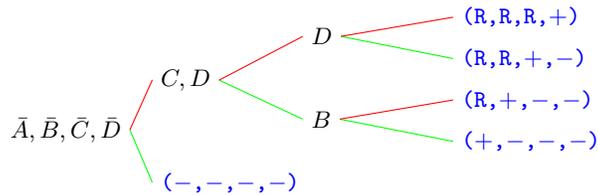

This algorithm performs an initial four-wise test, then adopts a divide-and-conquer strategy which we generalise below (in \Cref{sec:compound}), see \Cref{facto4}. Let us detail the operation of this algorithm.
If the first test is negative, we conclude that none of $A$, $B$, $C$ and $D$ are infected and proceed with a new set of four subjects.
If the first test is positive, we test $C$ and $D$ together. If this second test is positive, we conclude that $C$, $D$ or both are infected: we test $D$ and conclude as before. If the second test is negative, we conclude that $C$ and $D$ are not infected: we test $B$ and conclude as before.

\subsubsection{Algorithm $\mathcal A_5$.}
This algorithm is the most complex of the basic ones, and begins with a five-wise test, see \Cref{facto5}.

Let us consider we are testing individuals $A$, $B$, $C$, $D$ and $E$ together. If the first is negative, we conclude that none of the five subjects is infected and can restart the algorithm with a new set of individuals.
Otherwise, we test $A$ and $B$ together.

If the second test is positive, we re-inject $C$, $D$ and $E$ to the pool of individuals to be tested, test $B$ and conclude as before for $A$ and $B$. Otherwise, we conclude that at least one subject between $C$, $D$ and $E$ is infected. We pick two new subjects, say $F$ and $G$, and test $E$, $F$ and $G$ together. 

If the third test is negative, we conclude that $E$, $F$ and $G$ are not infected, and that $C$, $D$ or both are infected and conclude after testing $C$.

Otherwise, we then test $C$, $D$ and $G$ altogether. If this fourth test is negative, we conclude that $C$, $D$ and $G$ are not infected, implying that $E$ is infected (since $CDE$ was positive) 
Since $E$ is infected the $EFG$ test brings no information about $F$, which must be re-injected into the pool of subjects to be tested.

If the fourth test is positive, we then test $G$.

Otherwise, we conclude that $C$, $D$ or both are infected (since $CDG$ test was positive) and that $E$, $F$ or both are infected (since $EFG$ test was positive). We then test $D$ individually and $F$ individually and conclude as before.

We are left with the case where the test on $G$ alone is positive. While we easily conclude that $G$ is infected, we also conclude that the $EFG$ and $CDG$ tests do not bring any information about $C$, $D$, $E$ and $F$. It remains however the knowledge that $CDE$ had a positive test. We can thus go back to the position we were after the second test, and restart the process with a new individual $G'$ replacing $G$. 

\begin{remark}
The loopback in the process occurs rarely; the probability that this loop is taken several times is extremely small, however it is not zero. In practice, there may be a limit on the number of times a given individual can be tested. To avoid running in such issues it is possible to abort early by testing $C$ alone, or $D$ and $E$ together.
\end{remark}

\begin{figure}[!htp]
    \centering
    \scalebox{0.9}{
    \begin{tikzpicture}[grow=right,level distance=2.5cm,rotate=90]
\tikzset{every tree node/.style={anchor=base west}}
\Tree[.\node{$\overline{A},\overline{B},\overline{C},\overline{D},\overline{E}$} ;
    \edge[green] ; [.\node[blue]{\texttt{($-$,$-$,$-$,$-$,$-$)}} ; ] 
    \edge[red] ; [.\node{$A,B$} ;
        \edge[green] ; [.\node{\texttt{$E,\overline F, \overline G$}} ;
            \edge[green] ; [.\node{\texttt{$C$}} ;
                \edge[green] ; [.\node[blue]{\texttt{($-$,$-$,$-$,$+$,$-$,$-$,$-$)}} ; ]
                \edge[red]; [.\node[blue]{\texttt{($-$,$-$,$+$,R,$-$,$-$,$-$)}} ; ]
            ]
            \edge[red] ; [.\node{\texttt{$C,D,G$}} ;
                \edge[green] ; [.\node[blue]{\texttt{($-$,$-$,$-$,$-$,$+$,R,$-$)}} ; ]
       	        \edge[red] ; [.\node{\texttt{$G$}} ; 
                    \edge[green] ; [.\node{\texttt{$D$}} ;
                        \edge[green] ; [.\node{\texttt{$F$}} ;
                            \edge[green]; [.\node[blue]{\texttt{($-$,$-$,$+$,$-$,$+$,$-$,$-$)}} ; ] 
                            \edge[red]; [.\node[blue]{\texttt{($-$,$-$,$+$,$-$,R,$+$,$-$)}} ; ]
                        ]
                        \edge[red] ; [.\node{\texttt{$F$}} ;
                            \edge[green]; [.\node[blue]{\texttt{($-$,$-$,R,$+$,$+$,$-$,$-$)}} ; ] 
                            \edge[red]; [.\node[blue]{\texttt{($-$,$-$,R,$+$,R,$+$,$-$)}} ; ]
                        ]
                    ]
                    \edge[red] ; [.\node[red]{\texttt{L $E,F,\overline{H}$}} ; ]
                ]
           ]
         ]   
        \edge[red] ; [.\node{\texttt{$B$}} ;
                \edge[green] ; [.\node[blue]{\texttt{($+$,$-$,R,R,R)}} ; ]
                \edge[red] ; [.\node[blue]{\texttt{(R,$+$,R,R,R)}} ; ]
        ]
  ]
]
                
\end{tikzpicture}%
    }
    \caption{Algorithm $\mathcal A_5$.}\label{facto5}
\end{figure}

\subsection{Compound Algorithms}
\label{sec:compound}

Using the basic algorithms described in the previous section, we can build new algorithms as follows: choose an algorithm $\mathcal A_n$, and instead of applying algorithm $\mathcal A_n$ to individuals, we apply it on samples resulting from \emph{pairs} of individuals. The outputs of $\mathcal A_n$ will then need to be re-interpreted: a negative result means both members of the pair are negative, but a positive output for a mix $AB$ means that either $A$, $B$ or both are infected. As before, we test $B$. If test on $B$ is positive, $B$ is infected and we gained no information about $A$. If test on $B$ is negative, $B$ is not infected but $A$ is infected.

\begin{remark}
This generic construction yields $\mathcal A_2$ and $\mathcal A_4$ from $\mathcal A_1$ and $\mathcal A_2$ respectively. Therefore, these basic algorithms can be considered redundant.
\end{remark}
Starting with the sets of algorithms  \{$\mathcal A_1$,$\mathcal A_3$,$\mathcal A_5$\}, we get an infinite family of algorithms: \break
\begin{equation*}
    \mathcal A_1,\mathcal A_2,\mathcal A_3,\mathcal A_4,\mathcal A_5,\mathcal A_6,\mathcal A_8,\mathcal A_{10},\mathcal A_{12},\mathcal A_{16},\mathcal A_{20},\mathcal A_{24},\mathcal A_{32},\mathcal A_{40},\mathcal A_{48},\mathcal A_{64},\dotsc
\end{equation*}

\subsection{Complexity Analysis}\label{sec:complexityA}

Let $\mathcal A_n$ be one of the algorithms described above (basic or compound), we are interested in the number $f_n(x)$ which counts how many tests per person are needed on average to get the definitive status (positive or negative) of every individual, as a function of the population risk level $x$.

\paragraph{Algorithm $\mathcal A_1$.} We have, obviously, $f_1(x) = 1$. 

\paragraph{Algorithm $\mathcal A_3$.}
We first compute the probability of each leaf of the graph. For example, the leaf \textcolor{blue}{\texttt{($-$,$-$,$-$)}} is reached if, and only if, $A$, $B$ and $C$ are not infected, which has probability $(1-\rho)^3$. As a second example, the leaf \textcolor{blue}{\texttt{(R,R,$+$,$-$,$-$)}}  is reached when $C$ is infected, while $A$ and $B$ are not, disregarding the status of $A$ and $B$. As a result, the probability to reach this leaf of the graph is $\rho(1-\rho)^2$. Summing the number of tests needed to reach each leaf, weighted by the probability to reach this leaf, gives the average number of test per run of the algorithm. In the same vein, summing the 
number of known status, weighted by the probability to reach this leaf, gives the average number of patients whose status is discovered, per run of the algorithm. Dividing those two average numbers yields the desired value of average number of tests needed to get the status of one patient. For algorithm $\mathcal A_3$, the result is:
\begin{equation*}
f_3(x) = \frac{2x^4-6x^3+2x^2+6x+1}{x^3-3x^2+x+3}.
\end{equation*}

\paragraph{Algorithm $\mathcal A_5$.}
Analysis is similar to $\mathcal A_3$, with one additional complication: indeed, there is a possible loop back in the algorithm. A simple way to circumvent this is to expand the loop back and consider an infinite but rather simple graph and proceed as for $f_3$. The infinite series that appear have closed loop expressions (and are well known). We end up with:
\begin{equation*}
    f_5(x) = \frac{3x^6-18x^5+36x^4-24x^3-8x^2+13x+1}{(x^2-x-1)(x^3-5x^2+8x-5)}.
\end{equation*}

	
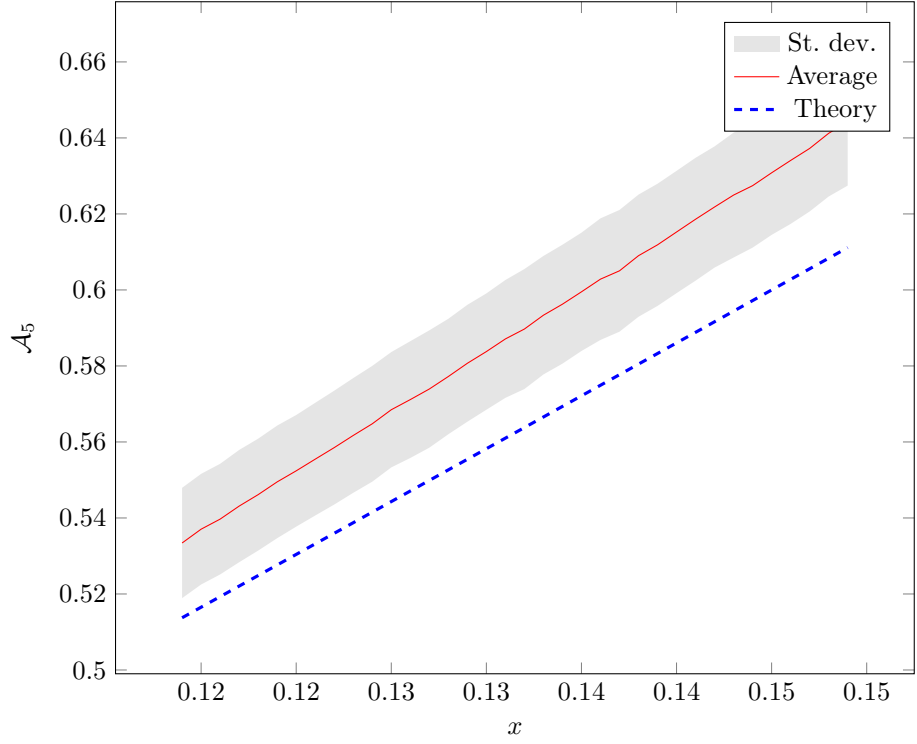
\begin{tikzpicture}
	\begin{axis}[legend cell align = right, width=\textwidth, legend pos = north east, ylabel= $\mathcal A_5$, xlabel = $x$]
   
	\addplot [stack plots=y, fill=none, draw=none, forget plot] 
	    table [x=p, y expr=\thisrow{avg}-\thisrow{std}] 
	    {data/a5.dat} \closedcycle;
	    
    \addplot [stack plots=y, fill=gray!50, opacity=0.4, draw opacity=0, area legend]
        table [x=p, y expr=\thisrow{std}+\thisrow{std}] 
        {data/a5.dat} \closedcycle;
    
    \addlegendentry{St. dev.};	
	
	\addplot [stack plots=false, thin, red]
	    table[x=p,y=avg]
	    {data/a5.dat};
	    \addlegendentry{Average};
	    
	\addplot [stack plots=false,very thick,smooth,blue,dashed]
	    table[x=p,y=theory]
	    {data/a5.dat};
	    \addlegendentry{Theory};	
	\end{axis}
	\end{tikzpicture}

\paragraph{Algorithms $\mathcal A_{2n}$.}
For compound algorithms, we can express $f_{2n}$ as a function of $f_n$. Let us denote $x_2$ the probability that at least one individual of a pair is infected, which is a simple function of $x$: $x_2 = 1-(1-x)^2 = 2x-x^2$.

The cost of the execution of $\mathcal A_n$ on a pair is on average $\alpha=f_n(x_2)$ to get the status of one pair. With probability $(1-x)^2$, we get the (negative) status of two patients at the cost of $\alpha$ tests on average. With probability $x(1-x)$, we get the (mixed) status of two patients at the cost of $\alpha+1$ tests in average. Finally, with probability $x$, we get the (positive) status of one patient at the cost of $\alpha$ tests in average. The average cost for one run is $x_2+f_n(x_2)$, while the average number of statuses determined in one run is $2-x$. We thus have:
\begin{equation*}
    f_{2n}(x) = \frac{x_2+f_n(x_2)}{2-x}. 
\end{equation*}
In particular, this allows us to express $f_2$ and $f_4$:
\begin{align*}
f_2(x) & = \frac{x^2-2x-1}{x-2}, &
f_4(x) & = \frac{2x^4-8x^3+12x^2-8x-1}{(x-2)(x^2-2x+2)}. 
\end{align*}

\subsection{Cut-off Points for Basic Algorithms}
Using the functions $f_n$ described above, we can identify which algorithm is the best at a given value of $x$. Because all the $f_n$ are rational functions in $x$, these regions of dominance are finite unions of intervals --- and in this particular case, they are simple intervals. In other terms, we can describe an algorithm's dominance region by specifying a \enquote{cutoff value} at which another algorithm becomes superior.

\begin{figure}[!ht]
    \centering
    \begin{tikzpicture}
    \begin{scope}[scale=2.0]
    
    \draw[thin,>=latex,->] (0, 0) to (5.5, 0); \node at (5.75, 0) {$x$};
    
    \draw[line width=1mm, blue] (1.14, 0) to (1.5, 0); \node[rotate=90,anchor=west,scale=0.9,blue] at (1.32, 0.125) {$\mathcal A_5$};
    \draw[line width=1mm, red] (1.5, 0) to (1.71, 0); \node[rotate=90,anchor=west,scale=0.9,red] at (1.605, 0.125) {$\mathcal A_4$};
    
    \draw[line width=1mm, blue] (1.71, 0) to (2.45, 0); \node[rotate=90,anchor=west,scale=0.9,blue] at (2.08, 0.125) {$\mathcal A_3$};
    
    \draw[line width=1mm, red] (2.45, 0) to (3.82, 0); \node[rotate=90,anchor=west,scale=0.9,red] at (3.135, 0.125) {$\mathcal A_2$};
    
    \draw[line width=1mm, blue] (3.82, 0) to (5, 0); \node[rotate=90,anchor=west,scale=0.9,blue] at (4.41, 0.125) {$\mathcal A_1$};
    
    \draw (0,-0.1) to (0, 0.1); \node at (0, -0.35) {$0$};
    \draw (5,-0.1) to (5, 0.1); \node at (5, -0.35) {$\frac12$};
    
    \draw (1.14,-0.1) to (1.14, 0.1); \node[scale=0.8] at (1.14, -0.35) {$\gamma_5$};
    \draw (1.5,-0.1) to (1.5, 0.1); \node[scale=0.8] at (1.5, -0.35) {$\gamma_4$};
    \draw (1.71,-0.1) to (1.71, 0.1); \node[scale=0.8] at (1.71, -0.35) {$\gamma_3$};
    \draw (2.45,-0.1) to (2.45, 0.1); \node[scale=0.8] at (2.45, -0.35) {$\gamma_2$};
    \draw (3.82,-0.1) to (3.82, 0.1); \node[scale=0.8] at (3.82, -0.35) {$\gamma_1$};
    
    \end{scope}
    \end{tikzpicture}
    \caption{Region in which each elementary algorithm reaches > 99\% optimality, as a function of probability $x$.}
    \label{fig:dominance}
\end{figure}
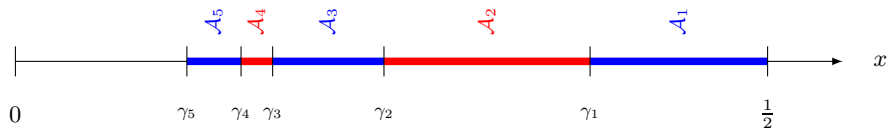

The value $\gamma_1$ is the one at which algorithms $\mathcal A_1$ and $\mathcal A_2$ have the same performances, i.e., it is a root of the numerator of the difference $f_1-f_2$. Therefore:
\begin{equation*}
    \gamma_1^2 -3\gamma_1 +1 = 0
\end{equation*}
Similarly, $\gamma_2$ is the cut-off point between $\mathcal A_2$ and $\mathcal A_3$, thus a root of $f_3(x)-f_2(x)$, which yields the equation:
\begin{equation*}
\gamma_2^3 -4\gamma_2^2 +5\gamma_2-1 = 0
\end{equation*}
Following this approach we obtain equations satisfied by all the cut-off points:
\begin{align*}
    \mathcal A_2/ \mathcal A_1 :{} & \gamma_1^2 -3\gamma_1 +1 = 0 \\
    \mathcal A_3/\mathcal A_2 :{} & \gamma_2^3 -4\gamma_2^2 +5\gamma_2-1 = 0 \\
    \mathcal A_4/\mathcal A_3 :{} & 2 \gamma_3^3 - 7 \gamma_3^2 + 7 \gamma_3 - 1 = 0 \\
    \mathcal A_5/\mathcal A_4 :{} & \gamma_4^9 - 10 \gamma_4^8 + 42 \gamma_4^7 - 96 \gamma_4^6 + 127 \gamma_4^5 - 91 \gamma_4^4 + 21 \gamma_4^3 + 14 \gamma_4^2 - 9 \gamma_4 + 1 = 0 \\
    \mathcal A_6/\mathcal A_5 :{} &\gamma_5^9 - 10 \gamma_5^8 + 44 \gamma_5^7 - 112 \gamma_5^6 + 179 \gamma_5^5 - 178 \gamma_5^4 + 98 \gamma_5^3 - 16 \gamma_5^2 - 8 \gamma_5 + 1 = 0
\end{align*}
which correspond to approximate values:
\begin{align*}
    \gamma_1 & = 0.381966011250105, &
    \gamma_2 & =0.245122333753307, &
    \gamma_3 & =0.170516459041503, \\ 
    \gamma_4 & =0.149636955876700, & 
    \gamma_5 & =0.113817389150325
\end{align*}
These values are illustrated on \Cref{fig:dominance}.

\Cref{fig:dominance} seems to point two shortcomings of basic algorithms: the region below $\gamma_5$ and the region above $x = 1/2$. For the former, we will discuss below how compound algorithms can provide a solution; for the latter, we formulate the following:
\begin{conjecture}
There is no better homogeneous population algorithm than $\mathcal A_1$ when $x > \gamma_1$: patients are to be tested individually. 
\end{conjecture}
Over their respective regions of dominance, we can compute the optimality of each algorithm with respect to the information-theoretical bound: $\mathcal A_1$ reaches 95.9\,\% for $\mathcal A_1$, and $\mathcal A_2$ through $\mathcal A_5$ all exceed 99\%.

Finally, for every $x < 0.23$, there exists $n$ and $k \in {1, 3, 5}$ such that $A_{2^nk}$ reaches 99\,\% optimality. Unfortunately, this fact does not help in selecting \emph{which} values of $n$ and $k$ to choose for a given value of $x$.

\section{Stratified Population Algorithms}
The results of the previous sections are very efficient for homogeneous populations with low risk level. However, they may be sub-optimal (by several percents) across some higher risk level ranges. 

This section addresses strategies consisting in mixing two groups. When facing more than two groups, the strategies described here can also be used on pairs of groups. Further algorithms can be derived based on the principles described in this section.

Once again, we focus on reaching the (arbitrary) minimal performance of 99\%. To avoid unnecessary complexity, we restrict ourselves to consider two populations, with risk levels $x$ and $y$ satisfying $x<y$ and $y<0.23$. This ensures that we already have at hand a quasi-optimal (i.e., performance above 99\%) algorithm for homogeneous populations with risk level $y$.

We also assume that the low risk population is much larger than the high-risk one. The strategy consists, therefore, of using a mix of subjects to deal with the high-risk ones. Then, we will be left with excess of low-risk patients, that we suggest to deal with as an homogeneous population.

\subsection{Basic Algorithms}
\subsubsection{Algorithm $\mathcal M_1$.}
This algorithm tests pairs of type $Ab$ with risk level $x$ for $A$ and risk level $y$ for $b$. Each time such test is negative, one concludes that $A$ and $b$ are not infected. Each time the test $Ab$ is positive, $b$ is sent to a pool of patients with probability $z=y/(x+y-xy)$. This second pool is tested using the best available algorithm for homogeneous population with risk level $z$. Then, as usual, if $b$ happens to be negative, we conclude that $A$ is positive and when b happens to be positive, we re-pool $A$.

\begin{figure}[!ht]
    \centering
    \begin{tikzpicture}[grow=right,level distance=2cm]
\Tree[
.\node{$\overline{A},\overline{b}$} ;
	\edge[green] ; [.\node[blue]{\texttt{($-$,$-$)}} ; ] 
	\edge[red] ; [.\node[purple]{$b$} ;
       	\edge[green] ; [.\node[blue]{\texttt{($+$,$-$)}} ; ]
       	\edge[red] ; [.\node[blue]{\texttt{(R,$+$)}} ; ]
        ]
	]
]
\end{tikzpicture}%
    \caption{Algorithm $\mathcal M_1$. Mixed pair test with re-pooling}\label{factoM1}
\end{figure}
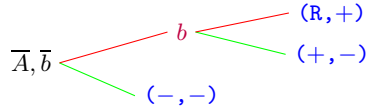

\Cref{factoM1} describes this algorithm, with test $b$ in purple to highlight it is not a direct test on a unique sample.

Let us note $\varphi(z)$ the cost function for best available algorithm for homogeneous population with risk level $z$. The function $\varphi$ can be picked amongst the cost functions detailed in \Cref{sec:complexityA}. The cost for execution of algorithm $M_1$ is then
\begin{equation*}
    1+(x+y-xy)\cdot\varphi\left(\frac{y}{x+y-xy}\right).
\end{equation*}
One execution of algorithm $\mathcal M_1$ brings surely knowledge about patient $b$'s status and brings knowledge about patient $a$ with probability $1-y$. As such, the overall performance of $\mathcal M_1$ is
\begin{equation*}
    \frac{(1-y)H(x)+H(y)}{1+(x+y-xy)\cdot\varphi(\frac{y}{x+y-xy})}
\end{equation*}

\subsubsection{Algorithm $\mathcal M_2$.}
\begin{figure}[!ht]
    \centering
    \begin{tikzpicture}[grow=right,level distance=2cm]
\Tree[
.\node{$\overline{A},\overline{b}$} ;
	\edge[green] ; [.\node[blue]{\texttt{($-$,$-$)}} ; ] 
	\edge[red] ; [.\node{$b,\overline{c}$} ;
       	\edge[green] ; [.\node[blue]{\texttt{($+$,$-$,$-$)}} ; ]
       	\edge[red]   ; [.\node{\texttt{$A$}} ; 
       	    \edge[green] ; [.\node[blue]{\texttt{($-$,$+$,R)}} ; ]
       	    \edge[red]   ; [.\node{\texttt{$b$}} ;
       	        \edge[green] ; [.\node[blue]{\texttt{($+$,$-$,$+$)}} ; ]
       	        \edge[red]   ; [.\node[blue]{\texttt{($+$,$+$,R)}} ;]
       	    ]
       	]
    ]
]
\end{tikzpicture}%
    \caption{Algorithm $\mathcal M_2$. Mixed pair test with re-pooling}\label{factoM2}
\end{figure}
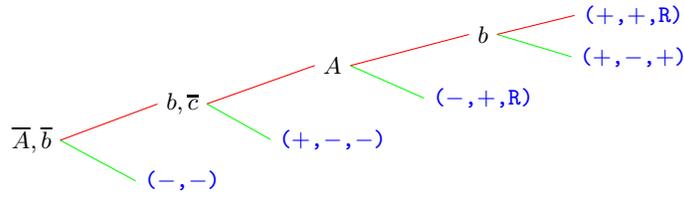

The algorithm $\mathcal M_2$ is described in \Cref{factoM2}. 

\subsubsection{Algorithm $\mathcal M_3$.}
\begin{figure}[!ht]
    \centering
    \begin{tikzpicture}[grow=right,level distance=2cm]
\Tree[
.\node{\texttt{$\overline{A},\overline{b}$}} ;
	\edge[red] ; [.\node      {\texttt{$b,\overline{c}$}} ;
	    \edge[red] ; [.\node {\texttt{$c$}} ; 
	        \edge[red] ; [.\node {\texttt{$b,\overline{d}$}} ;
	            \edge[red]   ; [.\node {\texttt{$d$}} ; 
	                \edge[red] ; [.\node       {$\cdots$} ; ]
	                \edge[green]   ; [.\node[blue] {\texttt{(R,$+$,$+$,$-$)}} ; ]
	            ]
	            \edge[green]   ; [.\node[blue] {\texttt{($+$,$-$,$+$,$-$)}} ; ]
	        ]
	        \edge[green]   ; [.\node[blue] {\texttt{(R,$+$,$-$)}} ; ]
	    ] 
	    \edge[green]   ; [.\node[blue] {\texttt{($+$,$-$,$-$)}} ; ] 
	]
	\edge[green] ;   [.\node[blue]{\texttt{($-$,$-$)}} ; ] 
]
\end{tikzpicture}%
    \caption{Algorithm $\mathcal M_3$. Mixed population recursive testing}.\label{factoM3}
\end{figure}
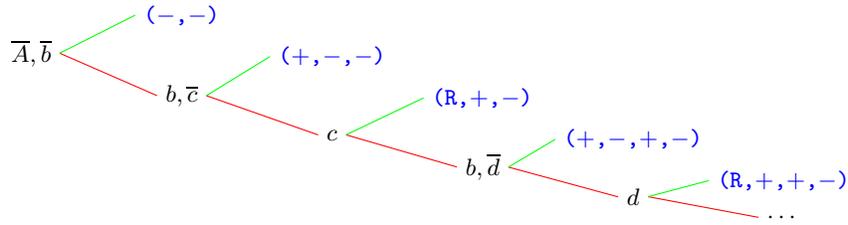

We consider a patient $A$ with risk level $x$ and the other patients with risk level $y$. We will start with testing $A$ and $b$ together. If the result is positive, we will determine the status of $b$, 

by testing $b$ with other patients with risk level $y$. See \Cref{factoM3}. The cost of running  the algorithm is once:
\begin{equation*}
    \frac{(1+y)(1+x-xy)}{1-y}
\end{equation*}
The algorithm provides knowledge about $A$ with probability $(1-y)$ and the average number of patients with risk level $y$ whose status is determined is:
\begin{equation*}
    \frac{(1+x-xy)}{1-y}
\end{equation*}
As a result, the performance of the algorithm, in terms of average information obtained per test is
\begin{equation*}
    \frac {(1-y)^2H(x)+(1+x-xy)H(y)} {(1+y)(1+x-xy)}  
\end{equation*}

\subsection{Combining Algorithms $\mathcal M_1$, $\mathcal M_2$ and $\mathcal M_3$}\label{sec:m123}
Depending on the values of $x$ and $y$, one will choose algorithm $\mathcal M_1$, $\mathcal M_2$, $\mathcal M_3$ or $\mathcal M_1$ applied to patients with risk level $x$ and pairs of patients with risk level $y$, followed by an additional test when a pair if found to be positive.

For all risk levels $31.25<x<\%,44.18\%$ and all risk levels $11\%<y<22\%$, a performance of at least 99\% can be reached by one of those four algorithms. For lower risk levels $y$, the same technique is applied using groups of $2^n$ patients. There always exist $n$ such that the probability of a group of $2^n$ has at least one of them infected falls in the range $31.25\%<x<,44.18\%$ and the tests needed to split the $2^n$ groups in $2^{n-1}$ have average entropy above $99\%$.

As a final result, for all risk levels $31.25\%<x<44.18\%$ and all risk levels $y\leq22\%$, we have built an algorithm that ensures at least $99\%$ optimality.

\subsection{Algorithms $\mathcal M_{4,n}$}
We will now define a family of algorithms that generalizes the algorithm $\mathcal M_1$. The algorithm $\mathcal M_{4,n}$ starts with testing one patient $A$ having risk level $x$ together with $n$ patients $b_i$ each having risk level $y$. If the test is negative, the $n+1$ patients are negative, and we are done.
In case the test is positive, we test $b_n$ then $b_{n-1}$ then $b_{n-2}$ and so on. Each of these patients is tested using one of the algorithm developed for homogeneous population, using the fact it has an \textit{a posteriori} probability $z_i$ which is a function of $x$ and $y$. If one of these test is positive, the patient $b_i$ is positive, patients $b_{i+1}$ to $b_n$ are negative, and we have to re-pool patients $A$ and $b_1$ to $b_{i+1}$. The remaining case is that all tests till $b_1$ are negative. We then conclude that $A$ is positive.


In \Cref{sec:m123} we addressed the \enquote{window} $x \in [0.3125,0.4418]$, where no known homogeneous algorithm reaches 99\% performance. The second \enquote{window} where we do not have a quasi-optimal algorithm (99\% performance) is the interval $[0.2345,0.25809]$. In this case, using $\mathcal M_{4,2}$, $\mathcal M_{4,3}$, $\mathcal M_{4,4}$ or $\mathcal M_{4,5}$, we can reach 99\% performance provided that the risk level $y$ is between 6\% and 18\% and the associated population is \enquote{large enough}. For cases where $y$ is lower than 6\%, we use recursively the same trick as before, creating a virtual population of risk level $y'= 2y-y^2$ by considering pairs of patients. When a pair is found positive, we test as usual on element of the pair and conclude as in algorithm $\mathcal M_2$.

As a final result, for all risk levels $23.45\%<x<,25.809\%$ and all risk levels $y\leq22\%$, we have built an algorithm that ensures at least $99\%$.

Combining with \Cref{sec:m123}, as soon as sufficiently large population with risk level $y$ below 18\% is available, we can manage a population with any risk level below 50\% with efficiency at least 99\%.

\section{Alternative Compound Strategies}
\label{sec:bizaro}

This part of the paper slightly improves performance, at the price of increasing the testing design complexity.

When drawing performance curves, one can notice that algorithm $\mathcal A_{16}$ seems inefficient compared to its neighbours, even if it fares better on a range of values for $x$. Instead of using $\mathcal A_{16}$, we consider $\mathcal A_{15}$: 
\begin{itemize}
    \item 
It starts with testing groups of 15 subjects. When the first test is positive, we test a subgroup of 6 subjects. We thus end up with a group of 6 or 9 subjects, at least one of which is infected. 

\item 
For subgroups of 6, we test the first one using one of the algorithm $\mathcal A_3$. If negative, we test the second one, again with algorithm $\mathcal A_3$. Either we identified an infected subject, or we are left with a subgroup of 4 subjects where one of which is infected. We perform two halving steps to conclude.

\item 
For subgroups of 9, we split them in subgroups of 4 or 5 by testing a set of 4 patients. Groups of 5 are again managed by testing one patient with $\mathcal A_3$. So either we are done, or we are again left with a group of 4 patients and we apply two halving steps.
\end{itemize}
Over the range of risk level $\rho$ where $\mathcal A_{16}$ outperforms $\mathcal A_{12}$ and $\mathcal A_{20}$, the new algorithm $\mathcal A_{15}$ outperforms $\mathcal A_{16}$. As a result, an improved sequence of algorithms is:
\begin{equation*}
\mathcal A_1, \mathcal A_2, \mathcal A_3, \mathcal A_4, \mathcal A_5, \mathcal A_6, \mathcal A_8, \mathcal A_{10}, \mathcal A_{12}, \mathcal A_{15}, A_{20}, \mathcal A_{24}, \mathcal A_{30}, \mathcal A_{40}, \mathcal A_{48}, \mathcal A_{60}, \dotsc 
\end{equation*}

Another construction allows for some optimization. The technique is to deal recursively with groups of 5 subjects. When the result is positive, one has to test the first subject of the group of 5, again using recursively previously known algorithm. If this individual is positive, one has to re-pool the other four. If this individual is not infected, we are left with a group of 4 subjects, of which at least one is infected. Two halving steps are performed to conclude.

Still another construction allows for some optimization. The technique is to deal recursively with groups of 9 subjects. When the result is positive, one has to test the first individual of the group of 9, again using recursively the previously known algorithm. If this individual is positive, one has to re-pool the other four. If the individual is not infected, we are left with a group of 8 subjects, one of which is infected. Three halving steps are performed to conclude.

\section{Open Questions}
Beyond the conjectures formulated in the course of this work, there are interesting questions left open for further research:
\begin{itemize}
    \item We assume perfect knowledge of the population risk $x$ to select the best algorithm. It seems that a slight error in the value of $x$ may in some situations cause us to select one of the neighbouring algorithms. Because the $f_n$ are rational (and hence continuous) this should have a limited effect, however this intuition should be formalised: what is the effect of having an uncertainty on $x$? 
    \item Our work builds on the assumption that dilution effects are negligible, and that tests are perfectly accurate. Lifting these hypotheses is left as a question for further research.
    \item Assume that there are two variants $V_1$ and $V_2$ of a disease, and we have tests $G$ (\enquote{generic}) and $S$ (\enquote{specific}), so that $G$ detects either of the variants and $S$ detects only one. What are the optimal strategies then to correctly identify individuals carrying $V_1$ and those carrying $V_2$? What if both variants can coexist?  
\end{itemize}

\bibliographystyle{abbrv}
\bibliography{biblio.bib}

\appendix

\end{document}